\documentclass[aps,prl,preprint,superscriptaddress, showpacs]{revtex4-1}

\usepackage{hyperref}
\usepackage{graphicx}
\begin{document}

%\preprint{}

\title{Spatiotemporal clustering and separation in regional earthquakes}


\author{Rene C. Batac}
\email[]{rbatac@pks.mpg.de}
%\thanks{}
\affiliation{Max-Planck-Institut f\"{u}r Physik komplexer Systeme, N\"{o}thnitzer Str. 38, 01187, Dresden, Germany}
\affiliation{National Institute of Physics, University of the Philippines Diliman 1101 Quezon City, Philippines}
\author{Holger Kantz}
\affiliation{Max-Planck-Institut f\"{u}r Physik komplexer Systeme, N\"{o}thnitzer Str. 38, 01187, Dresden, Germany}


\date{\today}

\begin{abstract}
Spatiotemporal clustering of earthquake events is a generally-established fact, and is important for designing models and assessment techniques in seismicity. Here, we investigate how this behavior can manifest in the statistical distributions of inter-event distances and times between earthquakes from different regional catalogs. We complement the analysis of previous authors (Touati \textit{et al.}, PRL 102, 168501 (2009)) and observe histograms best described by a superposition of two component distributions for ``short'' and ``long'' distances and times. Our results quantify the spatiotemporal clustering of earthquakes that are possibly generated by the same triggering mechanism. Independent earthquakes, on the other hand, are found to be separated by long inter-event distances and times. The statistics presented reveal regional differences, suggesting non-universality of the distributions.
\end{abstract}

%\keywords{}

\maketitle


Historical records of earthquakes allow us to deduce several underlying mechanisms of seismicity through the statistics of \textit{inter-event} properties, \textit{i.e.} the separation of successive earthquake events in space and in time. Many previous works have attempted to describe the form and implications of the distributions of these inter-event properties, using global and regional earthquake catalogs~\cite{DavidsenPaczuskiPRL2005,CorralPRL2006,BakEtAlPRL2002,CorralPRL2004,DavidsenGoltzGRL2004,TouatiEtAlPRL2009}. For example, in the Southern California seismic region, where extensive records exist for several decades of observation, earlier works report that both inter-event distances or ``jumps'' between earthquake epicenters ~\cite{DavidsenPaczuskiPRL2005,CorralPRL2006} and the inter-event times, or return times~\cite{BakEtAlPRL2002,CorralPRL2004}, exhibit statistical distributions involving power-law regimes, revealing the complex spatiotemporal (self-) organization of seismicity~\cite{SaichevSornettePRL2006}. 

Incidentally, we note that most authors focus on only one of these properties, \textit{i.e.} either the inter-event separation distance or return times. However, our understanding of the kinematics of earthquake formation suggests a direct relationship between these inter-event properties. While some authors observe universal return time distributions obtained upon rescaling the data~\cite{CorralPRL2004}, others argue that the spatial extent of observation plays a role in the distribution of return times~\cite{DavidsenGoltzGRL2004,TouatiEtAlPRL2009}. In particular, Touati \textit{et al.}~\cite{TouatiEtAlPRL2009} report observable differences between the return time distributions of regional and global earthquake catalogs: the histogram of inter-event times of Southern California earthquakes shows two distinct peaks, signifying the difference in characteristic waiting times between correlated (same aftershock sequence) and independent (different sequences) events, while global statistics reveals a single characteristic peak due to overlapping sequences from various locations. They explain the results using an epidemic-type aftershock sequence (ETAS) model~\cite{KaganKnopoffJGR1981,OgataJASA1988,SornetteHelmstetterPRL2002}, wherein the probability of generating a mainshock is used as proxy for spatial extent.

Here we aim to complement the analysis of Touati \textit{et al.}~\cite{TouatiEtAlPRL2009} by using an actual measure of epicenter separation distance in classifying the corresponding return time data. Our analyses are guided by the fact that spatiotemporal clustering is a well-established phenomenon in seismicity~\cite{Omori1894,KaganKnopoffGJRAS1980}, and must therefore manifest in any substantially large earthquake catalog regardless of region of origin and threshold magnitude. Therefore, instead of trying to find an approximate fitting function for the statistics of inter-event distances and times, we highlight the relationship between them using a simple procedure whose parameters are all derived from the data. Conditional distributions of earthquake return times subject to the corresponding spatial separation reveals that events separated by short (long) distances are also more likely to be separated by short (long) waiting times, clearly demonstrating clustering (separation) of correlated (independent) events. Interestingly, the different catalogs we used show a range of behaviors earlier observed in the ETAS model~\cite{TouatiEtAlPRL2009}, and suggest non-universal distributions.

We use three regional catalogs: the Philippines, PH (1973-2012), taken from the subset (4$^\circ$-24$^\circ$ N and 115$^\circ$-130$^\circ$ E) of the global Preliminary Determination of Epicenters catalog (PDE)~\cite{PDE19732012}; Japan, JP (1985-1998), from the Japan University Network Earthquake Catalog (JUNEC)~\cite{JUNEC19851998}; and Southern California, SC (1982-2012), from the Southern California Earthquake Database Center (SCEDC)~\cite{SCEDC19822012}. The PH catalog is relatively smaller with smallest recorded magnitudes of 2.7; the JP and SC catalogs, on the other hand, are large enough to allow for additional analyses for higher threshold magnitudes. The inter-event time $T$ between events $i$ and $i+1$ with magnitudes greater than or equal to a threshold magnitude $M$ is denoted by
\begin{equation}
\label{eqn:T}
T = t_{i+1} - t_{i}
\end{equation}

\noindent where $t$ denotes the actual time of occurrence. To complement the temporal analysis, the corresponding inter-event distance $R$ is defined as 
\begin{equation}
\label{eqn:R}
R = c R_E \sqrt{ \left( \phi_{i+1} - \phi_i \right)^2 + \left( \theta_{i+1} - \theta_i \right)^2}
\end{equation}

\noindent where the spatial coordinates are based on the longitude ($\phi$) and latitude ($\theta$) coordinates reported in degrees ($^\circ$) and converted to actual distance units by the scaling factor $c$ and the radius of the earth $R_E$, such that $1^\circ \approx 100$~km. This definition of the inter-event distance is a special case of the general hypocenter separation distance used by Kagan and Knopoff~\cite{KaganKnopoffGJRAS1980} and is valid only for regional catalogs, wherein $\Delta\phi, \Delta\theta \ll 1$. 

\begin{figure}
\centering
\includegraphics[width=\columnwidth]{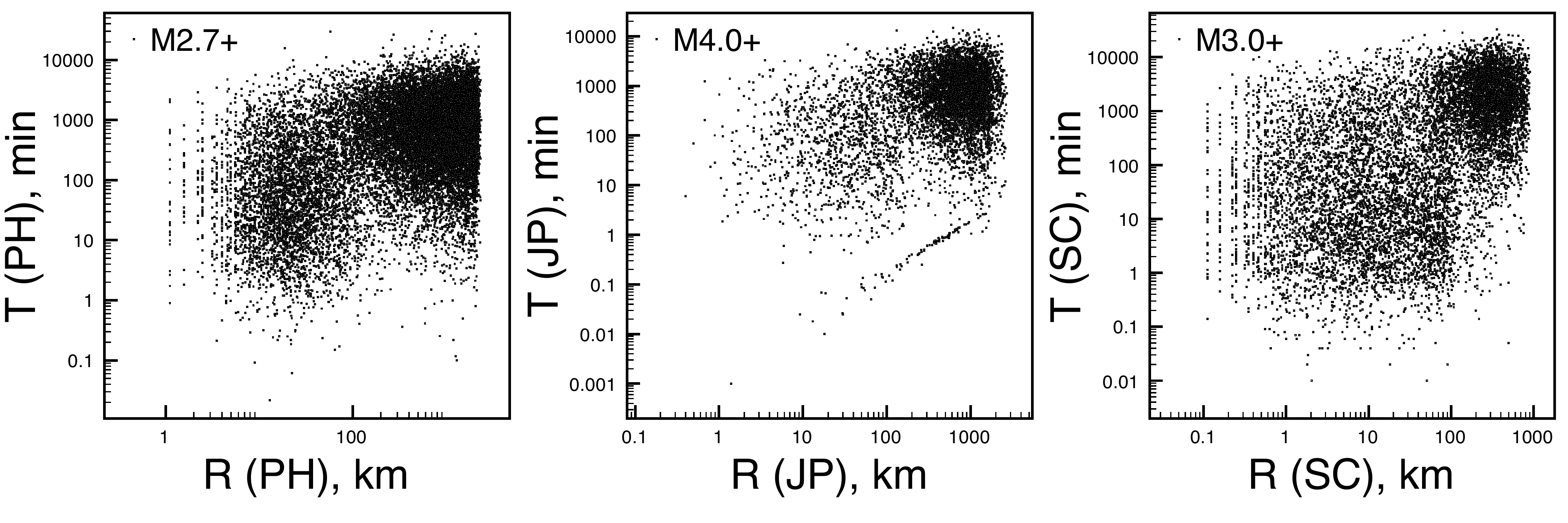}
\caption{\label{fig:RT} Scatter plots of all $R$ and $T$ values for the different regions considered. All plots show generally increasing trends, and reveal regimes of higher concentrations of points at the short $R$-$T$ and long $R$-$T$ regimes.}
\end{figure}

In Figure~\ref{fig:RT}, we plot all pairs of $R$ and $T$ values of single earthquake events in the regional catalogs (higher threshold magnitudes are presented for better visualization, but the behavior is the same for other threshold magnitudes). All scatter plots show a generally increasing trend, and closer inspection reveals dense concentration of points at both lower-left and upper right regimes of the scatter plots. To further highlight these two regimes, we present the distributions of $R$ and $T$ both in the form of unnormalized histograms $h(R)$ and $h(T)$ and normalized probability densities $p(R)$ and $p(T)$. All distributions are presented in logarithmic binning for better visualization in double-logarithmic plots. The inclusion of unnormalized histograms is made in view of previous observations that some features of the distributions may not be noticeable upon normalization~\cite{TouatiEtAlPRL2009}. 

\begin{figure}
\centering
\includegraphics[width=0.9\columnwidth]{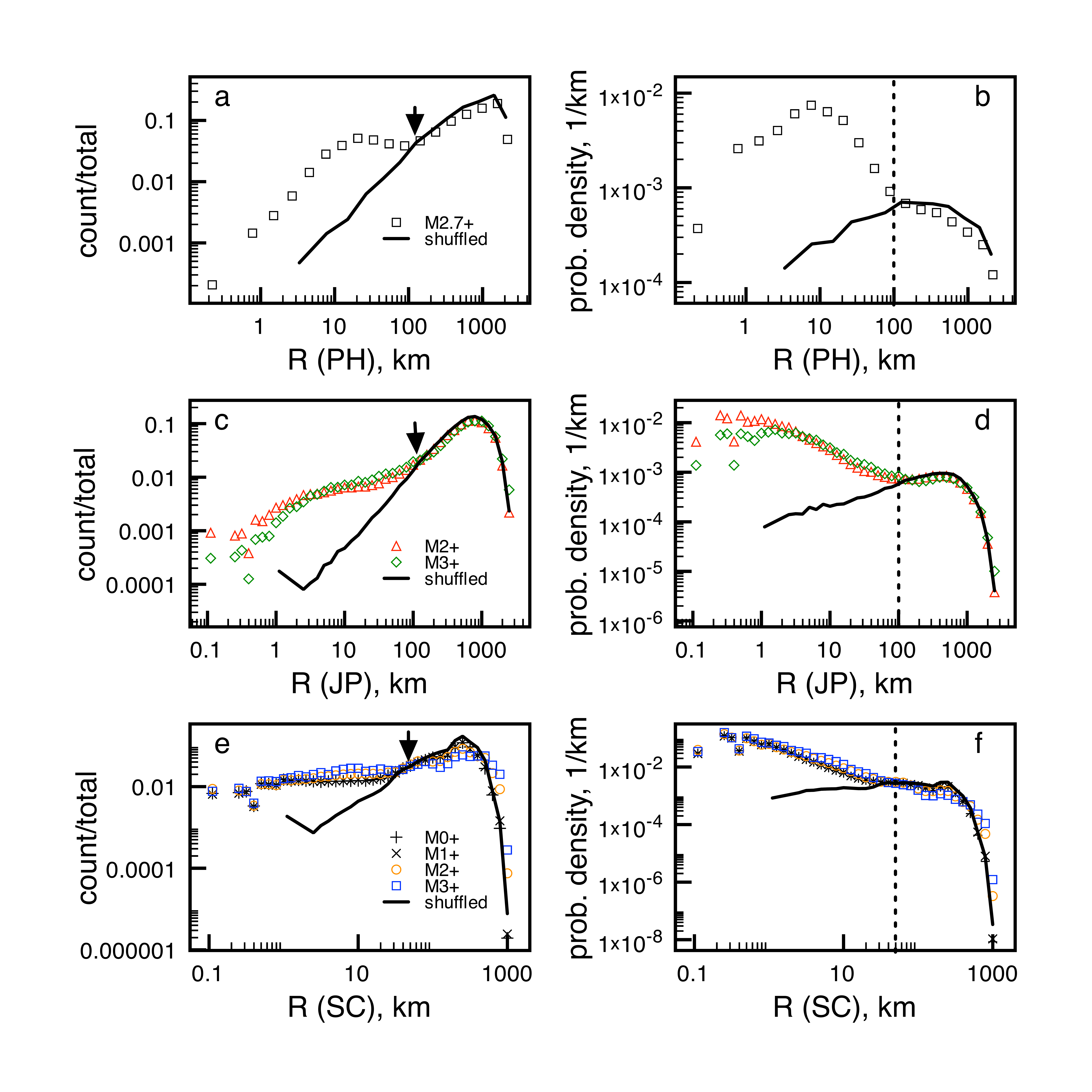}
\caption{\label{fig:R} (Color online.) Inter-event distance $R$ histograms (left) and probability densities (right) for the three data sets considered: (a)-(b) Philippines (events with magnitude 2.7 and above, M2.7+), (c)-(d) Japan (events M2.0+), and (e)-(f) Southern California (events M2.0+), superimposed with data obtained when the time series are randomly shuffled. The short-distance regimes from the original series are significantly of higher probability than the shuffled case, suggesting spatial clustering of correlated events. On the other hand, the original and shuffled series begin to follow the same trend beyond a certain $R^*$, denoted by the arrows in the left panels. These $R^*$ values are used to define the approximate boundaries for ``short'' and ``long'' distance events, shown by the dividing line in the right panels.}
\end{figure}

We observe two regimes in both $h(R)$ (left panels of Figure~\ref{fig:R}) and $p(R)$ (right panels of Figure~\ref{fig:R}), similar to those observed in a previous work~\cite{CorralPRL2006}. These two regimes are visually discernible from the shape of the distribution, and is consistent even for different threshold magnitudes. We interpret this as the superposition of two different distributions having different characteristic values. To approximate the boundary between the ``short'' and ``long'' $R$, we compare the original distributions with those generated from randomly-shuffled sequences. For each of the catalogs, we divide the entire time series into very short time slices of 0.1 s and labeled each slice by an index $n$ (for events that happened at exactly the same time, one of the events is moved to the next time slice). Randomly reordering the indices $n$ will therefore result in a random shuffling of the events~\cite{DavidsenPaczuskiPRL2005}. The histograms and probability densities of the randomly-shuffled $R$ are plotted as solid lines in Figure~\ref{fig:R}.

At short $R$ values, we observe that both the histograms $h(R)$ and probability density functions $p(R)$ of the original sequences are higher than those of the randomly shuffled series. Earthquakes occurring close together in space are more likely generated from the same aftershock sequence and are therefore correlated; expectedly, this correlated behavior is lost upon randomly shuffling the series, explaining the decline for short $R$ values in the distribution of shuffled events. However, after some value $R^*$ indicated by the arrows, the shape of the original and shuffled distributions begin to follow the same trend. This, on the other hand, can be attributed to the fact that earthquake events happening at very large distances away from each other are less likely to be correlated, and are thus generated by random, independent processes. We mark the following approximate values of $R^*$ for the different regions: 100 km for PH (Figure~\ref{fig:R}(a)); 100 km for JP (Figure~\ref{fig:R}(c)); and 50 km for SC (Figure~\ref{fig:R}(e)). The value of $R^*$ separates the ``short'' and ``long'' inter-event distances, as indicated by the broken lines in the right panels of Figure~\ref{fig:R}.

\begin{figure}
\centering
\includegraphics[width=0.9\columnwidth]{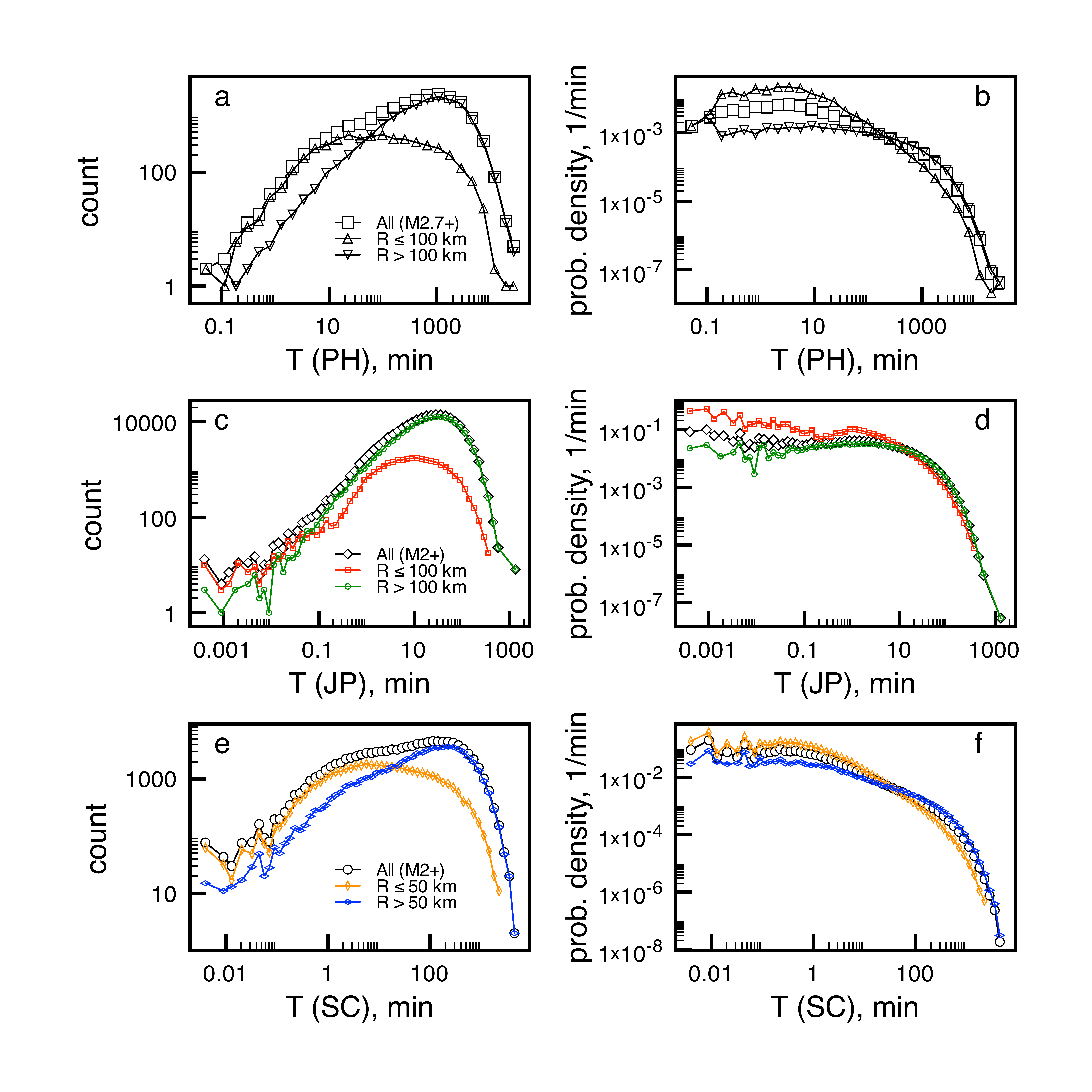}
\caption{\label{fig:T} (Color online.) Inter-event time $T$ histograms (left) and probability densities (right) for the three data sets considered: (a)-(b) Philippines (events with magnitude 2.7 and above, M2.7+), (c)-(d) Japan (events M2.0+), and (e)-(f) Southern California (events M2.0+). The conditional histograms and probability densities for the sets $T_{in} = \{T | R \leq R^*\}$ and $T_{out} = \{T | R > R^*\}$ are also shown. The histograms reveal various degrees of disparity between the time scales of short- and long-distance events, indicating the relative level of seismic activity in the regions. The probability densities show that events separated by short (long) distances are also more likely to be separated by short (long) time intervals.}
\end{figure}

This boundary, in turn, is used to separate the corresponding return times. We divided the set of all inter-event times into two groups based on the value of their corresponding $R$ relative to $R^*$: $T_{in} = \{T | R \leq R^*\}$ and $T_{out} = \{T | R > R^*\}$. Conditional distributions are important indicators of independence: if $T$ is independent of $R$, both conditional distributions of $T_{in}$ and $T_{out}$ should follow the same behavior, and collapse under the same curve upon normalization~\cite{LivinaEtAlPRL2005}. Our results, however, point to a strong dependence between these two properties. We present in Figure~\ref{fig:T} the conditional histograms, $h(T | R \leq R^*)$ and $h(T | R > R^*)$ plotted with the total histogram $h(T)$ (left panels) and the corresponding conditional density functions $p(T | R \leq R^*)$ and $p(T | R > R^*)$ and the total inter-event time probability density function $p(T)$ (right panels). 

The relationship between the total and conditional histograms in the left panels of Figure~\ref{fig:T} are reminiscent of the ETAS model results of Touati \textit{et al.}~\cite{TouatiEtAlPRL2009}. Using model-generated events, they observed that the total histogram results from the crossover of the distributions of correlated (same aftershock sequence) and independent (different aftershock sequence) events. In nature, however, it is nearly impossible to ascertain the origin of each individual seismic event to determine which ones are actually produced by the same mechanisms. Here, we show that grouping return times based on the corresponding inter-event distances results in a similar decomposition of the total distribution into two component distributions. Clustering in space and time is manifested by the fact that events happening at short $R$ have a $T$ distribution that tend toward shorter waiting times. On the other hand, we also observe a spatiotemporal separation of independent events, as events separated by long $R$ exhibit histograms that fit the tails of the $T$ distribution.

The same conclusion can be derived upon looking at the normalized density functions shown in the right panels of Figure~\ref{fig:T}. The component conditional density distributions do not collapse into the total probability density function. For the short-$T$ regime, we observe that $p(T | R > R^*) < p(T) < p(T | R \leq R^*)$, while for long-$T$, the reverse is true: $p(T | R \leq R^*) < p(T) < p(T | R > R^*)$. These inequalities confirm that short- (long-) distance events are more likely to happen within short (long) waiting times.

The difference between the conditional distributions and the total distribution can be viewed as a manifestation of the disparity in the time scales involved in the driving and relaxation mechanisms of earthquake events. The former, which we believe is responsible for the conditional histograms of long-range events, involves longer time scales, as it is driven by the slow process of tectonic motion (in the order of several cm/yr), and results in significantly long waiting times before the generation of a new independent event. The latter may explain the origin of shorter waiting time durations for nearby events, as individual earthquakes in the same aftershock sequence happen in minutes and entire sequences happen over a duration of several days or weeks. 

Interestingly, we believe that the resulting distributions can provide a hint as to the relative level of seismic activity in the region in consideration. This is inspired by the ETAS model results where the extent of disparity in the peaks of the component distributions are achieved by varying the rate of generation of mainshocks, $\mu$: Low-$\mu$ results in bimodal distributions (well-pronounced separation of the peaks of the component histograms) and high-$\mu$ results in unimodal distributions (overlapping peaks of component histograms)~\cite{TouatiEtAlPRL2009}. In the left panels of Figure~\ref{fig:T}, we observe that the disparity in the short characteristic peak of $h(T | R \leq R^*)$ and the long characteristic peak of $h(T | R > R^*)$ are very much pronounced for the case of the PH, Figure~\ref{fig:T}(a), and SC, Figure~\ref{fig:T}(c). On the other hand, the peaks of the conditional and total histograms for JP, Figure~\ref{fig:T}(b), show almost overlapping peaks, suggesting a relatively higher level of seismic activity in the region. In the end, the origin of these differences may ultimately be attributed to the differences in the fault properties; previous works, for example, have suggested similarities in the fault movements and structures of the Philippine and the San Andreas Faults~\cite{RutlandQJGS1967,AcharyaAggarwalJGR1980}, which are different from the highly complex fault networks found in Japan~\cite{MatsumotoEtAlGRL1992,HirataPAG1989}. 

%In addition, while previous works report this difference only between regional and global data sets~\cite{TouatiEtAlPRL2009}, our results show that this variability is also apparent for regional data sets, suggesting different mechanisms for different seismic regions, which can be linked with the physical properties (like fault properties and rate and type of crustal movement) of the regions under consideration.

%It is interesting to note that the differences between the component conditional distributions and the total distribution persists even upon normalization. Conditional densities are an important indicator of independence: if $T$ is independent of $R$, both conditional densities $p(T | R \leq R^*)$ and $p(T | R > R^*)$ are expected to collapse to the total probability density function $p(T)$~\cite{LivinaEtAlPRL2005}. However, as shown in the right panels of Figure~\ref{fig:T}, the component conditional density distributions $p(T | R \leq R^*)$ and $p(T | R > R^*)$ are not only different from $p(T)$, they also have different values for different return times. If we consider the short-$T$ regime for all regional data sets, we observe that $p(T | R > R^*) < p(T) < p(T | R \leq R^*)$. These relations suggest that nearby (far away) events are more (less) likely to happen shortly after one another in time. On the other hand, for long-$T$ regimes, the reverse is true: $p(T | R \leq R^*) < p(T) < p(T | R > R^*)$. This suggests relatively longer time separations between events happening at longer separation distances. 

\begin{figure}
\centering
\includegraphics[width=0.9\columnwidth]{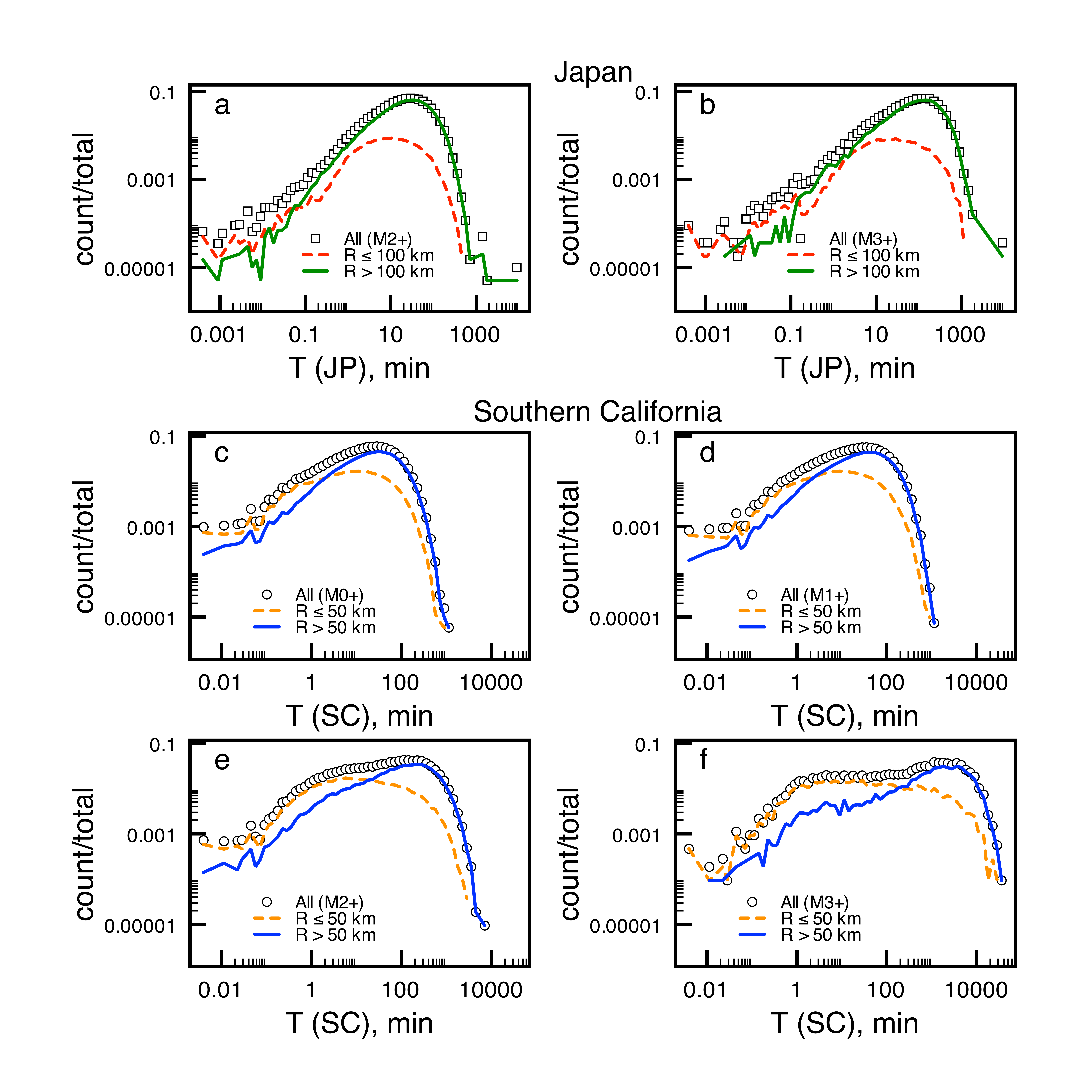}
\caption{\label{fig:TM} (Color online.) Conditional histograms (normalized with total number of events) $h(T | R \leq R^*)$ (broken lines) and $h(T | R > R^*)$ (solid lines) plotted alongside the histograms of all events $h(T)$ (symbols) for different catalogs and threshold magnitudes $M$: Japan, for (a) $M = 2$ and (b) $M = 3$; and Southern California, for (c) $M = 0$; (d) $M = 1$; (e) $M = 2$; and (f) $M = 3$. The scaling of the axes are preserved for (a)-(b) and for (c)-(f) for easier comparison. Aside from the expected lengthening of the return times for higher threshold magnitudes, we observe a widening gap between the peak values of the component histograms.}
\end{figure}

Finally, the spatio-temporal clustering of correlated events and the separation of independent events is still observed even for higher threshold magnitudes. Upon considering higher threshold magnitudes, weaker events are neglected from the analysis, thereby lengthening the mean waiting time between the occurrence of two ``successive'' events. Despite this, we still observe the separation of the temporal histogram into two conditional histograms based on separation distance. Using the same $R^*$ values shown in Figure~\ref{fig:R}, we analyze the JP and SC catalogs for different threshold magnitudes up to $M = 3$. The results of are presented in Figure~\ref{fig:TM}. Interestingly, despite the broadening of the distribution, the peak values of $h(T | R < R^*)$ for different threshold magnitudes roughly coincide at around 10 min for both JP and SC, hinting at the same correlated mechanisms. On the other hand, the peak of the $h(T | R > R^*)$ shifts to longer $T$ values for higher threshold magnitudes, suggesting that the broadening of the distribution for higher threshold magnitudes is primarily a result of the lengthening of the waiting time between independently generated events.

In summary, analysis of the inter-event properties of earthquakes from several regional datasets has allowed us to observe both spatiotemporal clustering and separation between successive earthquake events. Apart from showing that both the inter-event distances and times between successive earthquakes show two regimes that may be attributed to the nature of their triggering, we have described their relationship through analysis of conditional distributions. Events happening at close proximity to each other are shown to be more likely to happen after shorter waiting times, demonstrating the clustering behavior of correlated earthquakes simultaneously in space and time. We have also shown that events happening at long separation distances are more likely to also be separated by long waiting times, which, in turn, hints at independent mechanisms generating these events. Because of the inherent difficulty in accurately identifying the actual origins of individual earthquake events, classifying ``short'' and ``long'' separation distances from the distribution of inter-event distances provides a good approximation of the extent of the spatiotemporal clustering behavior. Finally, due to the differences observed in the distributions obtained for the different regions, we believe that the obtained inter-event distributions are non-universal and are highly affected by the local earthquake generating mechanisms.

\bibliography{Maintext}

\end{document}